\documentclass[11pt,a4paper]{article}

\usepackage[a4paper, total={18cm, 26cm}]{geometry}

\usepackage[latin1]{inputenc}
\usepackage[T1]{fontenc}

\usepackage{graphicx}

\usepackage{amsmath}
\usepackage{amssymb}

\usepackage[footnotesize, nooneline]{caption}

\newenvironment{sciabstract}
{\begin{quote} \bf}
{\end{quote}}

\newcommand{\ket}[1]{\ensuremath{\left| #1 \right\rangle}}

\title{Teleportation of entanglement over 143 km}

\usepackage{authblk}

\author[1,2]{Thomas Herbst\thanks{Correspondence: thomas.herbst@univie.ac.at}}
\author[1]{Thomas Scheidl}
\author[1]{Matthias Fink}
\author[1]{Johannes Handsteiner}
\author[1,2]{Bernhard Wittmann}
\author[1,2]{Rupert Ursin}
\author[1,2]{Anton Zeilinger}
\affil[1]{Institute for Quantum Optics and Quantum Information (IQOQI), Austrian Academy of Sciences, Boltzmanngasse 3, A-1090 Vienna, Austria}
\affil[2]{Vienna Center for Quantum Science and Technology, Faculty of Physics, University of Vienna, Boltzmanngasse 5, A-1090 Vienna, Austria}
\affil[  ]{}
\affil[ ]{(Dated: \today)}
\date{}

\usepackage{caption}
\begin{document}

\maketitle

\begin{sciabstract}
As a direct consequence of the no-cloning theorem \cite{Wootters:1982wj}, the deterministic amplification as in classical communication is impossible for quantum states. This calls for more advanced techniques in a future global quantum network \cite{Bose:1998wq, Kimble:2008if}, e.g. for cloud quantum computing \cite{Cirac:1999tz, Nielsen:2000vn, Ladd:2010kq}. A unique solution is the teleportation of an entangled state, i.e. entanglement swapping \cite{Zukowski:1993us}, representing the central resource to relay entanglement between distant nodes. Together with entanglement purification \cite{Bennett:1996wz, Briegel:1998wp, Pan:2001gg, Pan:2003kv} and a quantum memory \cite{Clausen:2011bw, Gisin:2007vl} it constitutes a so-called quantum repeater \cite{Briegel:1998wp, Duan:2001tt}. Since the afore mentioned building blocks have been individually demonstrated in laboratory setups only, the applicability of the required technology in real-world scenarios remained to be proven. Here we present a free-space entanglement-swapping experiment between the Canary Islands of La Palma and Tenerife, verifying the presence of quantum entanglement between two previously independent photons separated by 143 km. We obtained an expectation value for the entanglement-witness operator, more than 6 standard deviations beyond the classical limit. By consecutive generation of the two required photon pairs and space-like separation of the relevant measurement events, we also showed the feasibility of the swapping protocol in a long-distance scenario, where the independence of the nodes is highly demanded. Since our results already allow for efficient implementation of entanglement purification, we anticipate our assay to lay the ground for a fully-fledged quantum repeater over a realistic high-loss and even turbulent quantum channel.
\end{sciabstract}
\subsection*{Introduction}
The vision of a global quantum internet is to provide unconditionally secure communication \cite{Nielsen:2000vn, Gisin:2007vl}, blind cloud computing \cite{Barz:2012uw} and an exponential speedup in distributed quantum computation \cite{Nielsen:2000vn, Ladd:2010kq}. Teleportation of an entangled state, also known as entanglement swapping, plays a vital role in this vision. To date the entanglement-swapping protocol has been implemented in many different systems \cite{Pan:1998tn, Jennewein:2001vs, Halder:2007eo, Yuan:2008fj, Kaltenbaek:2009ix, Ma:2012kl, Hofmann:2012jb}, owing to the fact that it represents a key resource for numerous quantum-information applications. Since an unknown single quantum state cannot be cloned nor amplified without destroying its essential quantum feature, the quantum repeater is the main resource for faithful entanglement distribution over long distances. The idea is to decompose the total distance into shorter elementary links, over which entanglement is shared, purified and eventually stored in quantum memories from which the entangled states can be retrieved on demand, once all the nodes are readily prepared. Finally the entanglement is swapped between adjacent nodes and faithfully extended over the whole distance. Entanglement purification and quantum memories serve solely to enhance the efficiency and the fidelity of the protocol, both of which are limited due to imperfection of the sources of entangled particles, of the involved quantum operations and of the interconnecting quantum channels. Entanglement swapping however provides the underlying non-classical correlations and constitutes the fundamental resource required for the implementation of a quantum repeater. Here we show that we were able to provide this resource via a realistic 143 km long-distance free-space (elementary) link under harsh atmospheric conditions, representing the largest geographical separation for this protocol to date. Furthermore, the simultaneous creation of two randomly generated photon pairs drastically reduces the signal-to-noise ratio, leading to technological requirements on the verge of practicability. Nonetheless, we ensured space-like separation of the remote measurement events, which is important for certain protocols e.g. quantum key distribution \cite{Ekert:1991zz, Scheidl:2010fj}.
\\
The entanglement swapping protocol is realized via the generation of two entangled pairs, photons "0" and "1" and photons "2" and "3", for example the maximally entangled singlet states
\begin{align}
    &\ket{\Psi^-}_{01}=\frac{1}{\sqrt{2}}(\ket{H}_0\ket{V}_1-\ket{V}_0\ket{H}_1)\nonumber\\
    &\ket{\Psi^-}_{23}=\frac{1}{\sqrt{2}}(\ket{H}_2\ket{V}_3-\ket{V}_2\ket{H}_3)
\end{align}
where $\ket{H}$ and $\ket{V}$ denote the horizontal and vertical polarization states, respectively. The product state $\ket{\Psi}_{0123} =\ket{\Psi^-}_{01}\otimes\ket{\Psi^-}_{23}$ may be written as
\begin{align}
    \ket{\Psi}_{0123}=\frac{1}{2}[&\ket{\Psi^+}_{03}\otimes\ket{\Psi^+}_{12}-\ket{\Psi^-}_{03}\otimes\ket{\Psi^-}_{12}\nonumber\\
    -&\ket{\Phi^+}_{03}\otimes\ket{\Phi^+}_{12}+\ket{\Phi^-}_{03}\otimes\ket{\Phi^-}_{12}].
\end{align}
Therefore a so-called Bell-state measurement (BSM) between photons "1" and "2" results randomly in one of the four maximally entangled Bell states $\ket{\Psi^\pm}_{12}=\frac{1}{\sqrt{2}}(\ket{H}_1\ket{V}_2\pm\ket{V}_1\ket{H}_2)$ and $\ket{\Phi^\pm}_{12}=\frac{1}{\sqrt{2}}(\ket{H}_1\ket{H}_2\pm\ket{V}_1\ket{V}_2)$ with an equal probability of 1/4. By that measurement, photons "0" and "3" are projected into the same entangled state as photons "1" and "2". Thus the entanglement is swapped from photons "0-1" and "2-3" to the photons "1-2" and "0-3" (Figure \ref{figure1}a). This procedure may also be seen as teleportation of the state of photon "1" onto photon "3" or photon "2" onto photon "0". Although the implementation of this protocol, solely based on linear optics, allows distinguishing between two out of four Bell states only \cite{Calsamiglia:2001tl}, it provides a maximal fidelity of 1.

\subsection*{Experiment}
\begin{figure}[!t]
  \centering
      \includegraphics[width=0.8\textwidth]{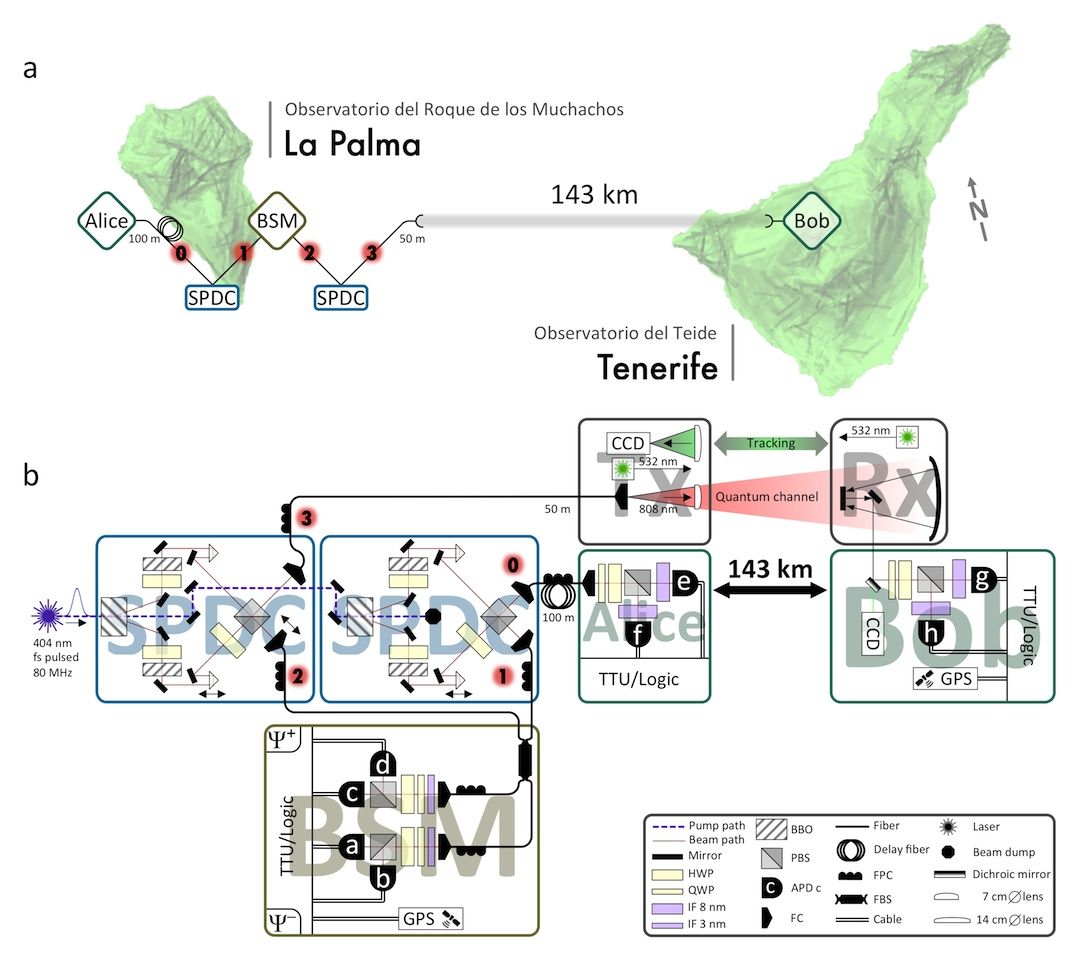}
  \caption[]{\textbf{Entanglement swapping over a 143 km free-space channel between the Canary Islands La Palma and Tenerife. a, }\textit{Experimental scheme}. Both spontaneous parametric down-conversion (SPDC) sources, the Bell-state measurement (BSM) module and Alice were situated on La Palma and Bob on Tenerife. The two SPDC sources generated the entangled photon pairs "0-1" and "2-3". Photons "1" and "2" (photons are indicated by black numbers on red circles) were subjected to a BSM. A 100 m fibre delayed photon "0" with respect to photon "3", such that Alice's and Bob's measurements were space-like separated. Revealing entanglement of photons "0" and "3" between Alice and Bob verified successful entanglement swapping. \textbf{b, }\textit{Experimental setup}. Two polarization-entangled photon pairs $\ket{\Psi^-}_{01}$ and $\ket{\Psi^-}_{23}$ were generated in two identical sources via spontaneous parametric down-conversion (SPDC) in a nonlinear $\beta$-barium borate (BBO) crystal. The photons were then coupled into single-mode (SM) fibres with fibre couplers (FC). Any polarization rotation in the SM fibres was compensated for by fibre polarization controllers (FPC). Photons "1" and "2" were spectrally filtered with interference filters (IF) with a full width at half maximum (FWHM) of 3 nm and overlapped in a fibre beam splitter (FBS). A subsequent polarization-dependent measurement was performed, utilizing a quarter wave plate (QWP), a half wave plate (HWP), a polarizing beam splitter (PBS) and four avalanche photodiodes (APD) a,b,c and d in the Bell-state measurement (BSM). Photon "3" was guided via a 50 m fibre to the transmitter (Tx) and sent to Bob in Tenerife, while photon "0" was delayed by a 100 m fibre before its polarization detection at Alice. The receiver (Rx) on Tenerife captured photon "3" where Bob performed his polarization-dependent measurement. Both Alice and Bob spectrally filtered their photons with IFs with 8 nm FWHM. All detection events were time stamped by time-tagging units (TTU) with a resolution of 156 ps and stored for subsequent analysis. See main text for detail.
}
  \label{figure1}
\end{figure}

Here we report successful entanglement swapping in an experiment performed on the Canary Islands, utilizing a 143 km horizontal free-space link between the Jacobus Kapteyn Telescope (JKT) building of the Isaac Newton Group of Telescopes (ING) on La Palma and the Optical Ground Station (OGS) of the European Space Agency (ESA) on Tenerife (Figure \ref{figure1}a). Both buildings are located at an altitude of 2400 m. The JKT served as the base station for the production of the two entangled photon pairs, for the BSM between photons "1" and "2" and for the polarization detection of photon "0" at Alice. The transmitter telescope, sending photon "3" to the receiving station on Tenerife, was installed on the rooftop of the JKT building. At the receiver the photons were collected by the 1 m diameter OGS reflector telescope and guided through the optical Coudé path to the setup for polarization analysis and the final measurement by Bob.
\\
In our experimental setup (Figure \ref{figure1}b) a mode-locked femtosecond pulsed Ti:Saph laser emitted light with a central wavelength of 808 nm at a repetition rate of 80 MHz. Those near-infrared pulses were then frequency doubled to a central wavelength of 404 nm using second-harmonic generation in a type-I nonlinear $\beta$-barium borate (BBO) crystal. The individual polarization-entangled photon pairs employed in the protocol were generated via spontaneous parametric down-conversion (SPDC) in two subsequent type-II phase-matched BBOs \cite{Kwiat:1995ub} and coupled into single-mode (SM) optical fibres for spatial mode selection. The quality of entanglement was optimized by eliminating the spectral distinguishability \cite{Kim:2002gd,Kim:2003tva,Poh:2009iq}, which is inherent to pulsed SPDC schemes. The first SPDC source provided the entangled state $\ket{\Psi^-}_{23}$, with photon "2" as one input photon for the BSM and photon "3" being guided through a 50 m long SM fibre to the transmitter telescope.  The second SPDC source prepared the state $\ket{\Psi^-}_{01}$, where photon "1" was the second input photon for the BSM. Photon "0" was locally delayed in a 100 m fibre ($\sim500$ ns) and subsequently measured by Alice, thus ensuring space-like separation between Alice's and Bob's measurement events \cite{Scheidl:2010fj}.
\\
In La Palma, the BSM was implemented using a tuneable fibre beam splitter (FBS) set to a 50:50 splitting ratio. While the spatial overlap of photons "1" and "2" is inherent to the FBS, a perfect temporal overlap is accomplished in the minimum of the Hong-Ou-Mandel \cite{Hong:1987vi} (HOM) interference dip. The latter was achieved by adjusting the optical path length for photon "2", by linearly moving the SM fibre coupler in the first SPDC source. Both output arms of the FBS were equipped with a quarter- and a half-wave plate followed by a polarizing beam splitter (PBS) in order to project on any desired polarization measurement basis. Intrinsic polarization rotations in the SM fibres were compensated for with in-fibre polarization controllers. The avalanche photodiodes (APDs) a, b, c and d, placed at the four outputs of the two PBSs, were connected to a home-made coincidence logic, providing the two valid outcomes of our BSM: simultaneous clicks of APDs (a $\&$ d) $\vee$ (b $\&$ c) or (a $\&$ b) $\vee$ (c $\&$ d) indicated that photons "1" and "2" were projected onto the maximally entangled $\ket{\Psi^-}_{12}$ singlet or $\ket{\Psi^+}_{12}$ triplet Bell state, respectively. As can be seen from Eq.(2), conditioned on these BSM results, photons "0" and "3" were thus simultaneously projected onto the very same states $\ket{\Psi^-}_{03}$ and $\ket{\Psi^+}_{03}$, respectively. The projection onto the other Bell-states $\ket{\Phi^\pm}_{12}$ does not result in a coincidence detection event by the BSM and thus cannot be resolved with a linear-optics scheme. Furthermore, the two valid BSM outcomes together with Alice's detection events of photon "0" (APDs e and f) were fed into a logic AND gate, providing four possible combinations. These local 3-fold coincidence events on La Palma as well as the remote detection events of photon "3" on Tenerife (APDs g and h) were then recorded by two separate time-tagging units (TTUs) with a temporal resolution of 156 ps. In order to retrieve the final 4-folds between Alice's events and those measured on Bob's side we calculated the cross-correlation between the remotely recorded individual measurement data - both synchronized to GPS standard time. To compensate for residual relative clock drifts between the distant TTU-clocks we employed entanglement assisted clock synchronization \cite{Ma:2012ei} between consecutive 30 s measurements, allowing for a coincidence-time window of down to 5 ns.
\begin{figure}[!b]
  \centering
    \includegraphics[width=0.75\textwidth]{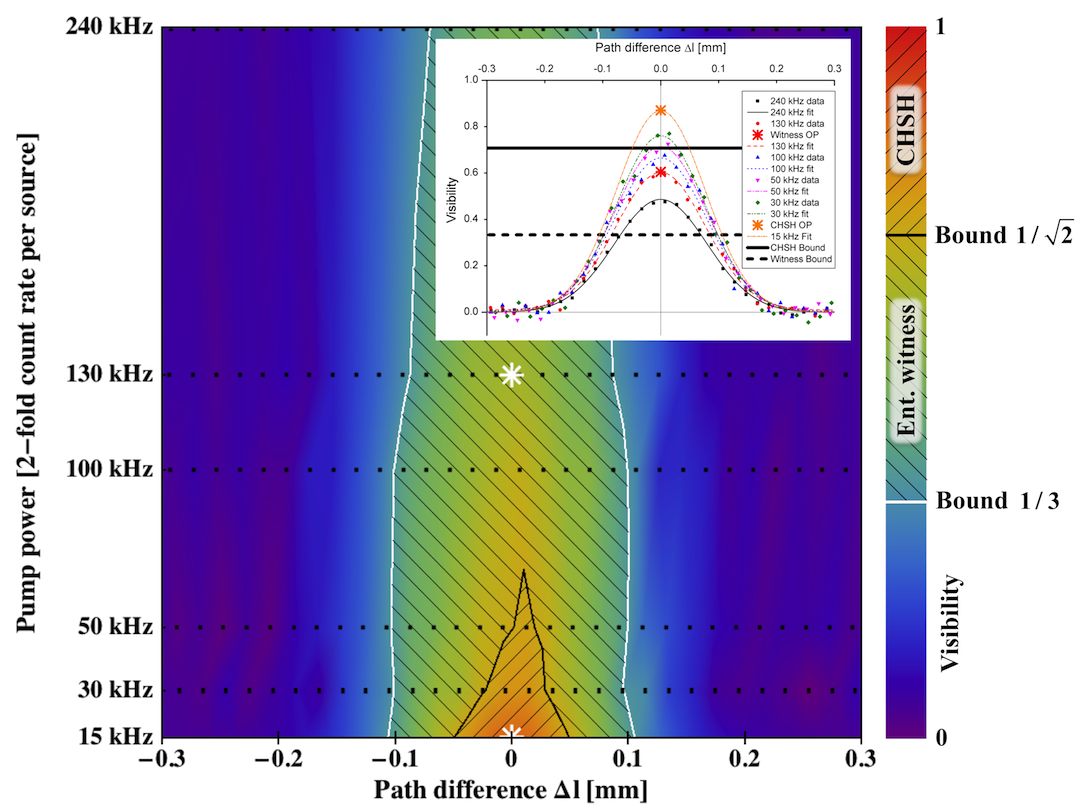}
  \caption[]{\textbf{A density plot of the entanglement-swapping visibility versus path-length difference and pump power measured locally on La Palma.} The abscissa represents the relative optical path-length difference $\Delta l$ between photon "1" and "2" in the Bell-state measurement and the ordinate represents the 2-fold count rate of the sources. All measured data points are indicated by black dots in the density plot. The entanglement-swapping visibility is illustrated by rainbow colours where the entanglement-witness bound and the bound for the violation of a Clauser-Horne-Shimony-Holt (CHSH) type Bell inequality are represented by a white and a black isoline at a visibility of 1/3 and $1/\sqrt{2}$, respectively. The hatched areas between visibilities of 1/3 and $1/\sqrt{2}$ and above $1/\sqrt{2}$ mark the regions where entanglement can be shown by an entanglement witness and a CHSH type Bell inequality, respectively. A plot of the individual measurement runs including the respective Gaussian fits is illustrated in the inset. The operating point (OP) for the local test of the CHSH inequality (see orange star in inset) was chosen at a 2-fold rate of 15 kHz and zero delay. At this set point a visibility of 0.87 was achieved. For the entanglement-swapping experiment via the 143 km and -32 dB free-space link we tuned the setup to a 2-fold rate of 130 kHz and again perfect overlap at zero delay, resulting in a visibility of 0.6 (see red star in inset). In the density plot white stars indicate both OPs.
}
  \label{figure2}
\end{figure}
\subsection*{Experimental results}
\begin{figure}[!b]
  \centering
    \includegraphics[width=0.7\textwidth]{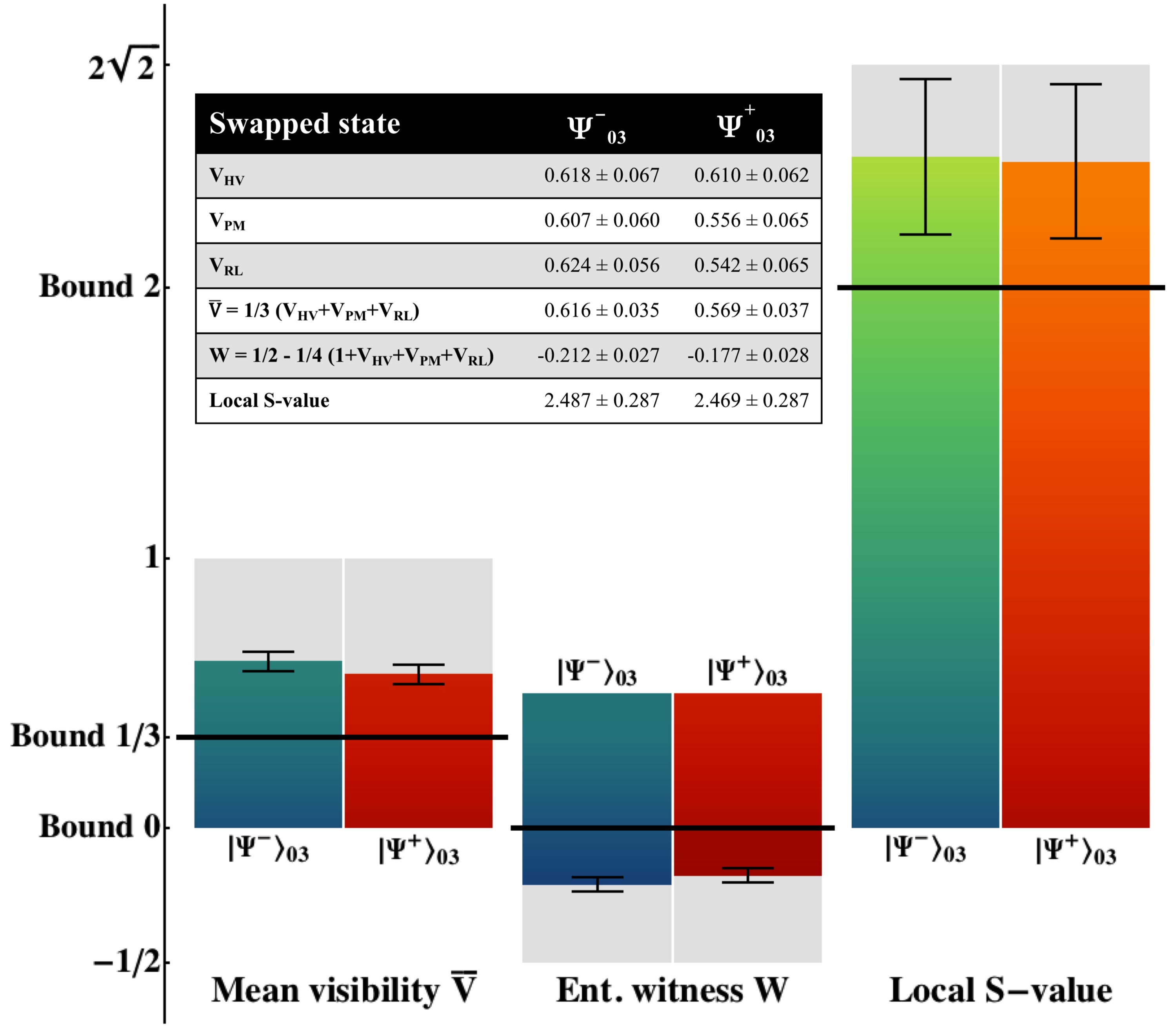}
  \caption[]{\textbf{Summary of the measurement results.} Over the 143 km free-space link we obtained an expectation value for the entanglement-witness operator $W$, more than $6\sigma$ below the classical bound of 0. This proved the presence of entanglement between the states $\ket{\Psi^-}_{03}$ and $\ket{\Psi^+}_{03}$. The violation of a Clauser-Horne-Shimony-Holt (CHSH) type Bell inequality was shown locally on La Palma underlining the quality of our setup. The bar chart illustrates the mean visibility $\overline{V}$ and the entanglement-witness operator $W$ over the 143 km link as well as the locally measured $S$-value. All outcomes are given for the swapped states $\ket{\Psi^-}_{03}$ and $\ket{\Psi^+}_{03}$ with error bars derived from Poissonian statistics, $\pm1\sigma$. The numerical values of the measurement results are given in the inset, including the individual visibilities $V_{HV}, V_{PM}$ and $V_{RL}$ in the three mutually unbiased bases horizontal/vertical (HV), plus/minus (PM) and right/left (RL) as well as the mean visibility $\overline{V}$, the entanglement-witness operator $W$ and the locally measured $S$-value.
}
  \label{figure3}
\end{figure}

The strong average attenuation of -32 dB over the 143 km free-space quantum channel calls for high production rates of the SPDC sources in order to operate well above the noise level of the single-photon detectors on Tenerife. However, pumping the SPDC sources with high pump intensities reduces the achievable entanglement-swapping visibility due to increased multi-pair emissions. Hence, a reasonable trade-off between count rates and visibility was required. In order to find the optimal operating point, we locally characterized our setup for various 2-fold count rates of the SPDC sources (Figure \ref{figure2}). The entanglement-swapping visibilities of our setup varied between 0.87 at lowest 2-fold count rate (15 kHz 2-folds, about 1 Hz 4-folds) and 0.49 at full pump power (240 kHz 2-folds, 370 Hz 4-folds).
\\
A traditional measure of entanglement is constituted by violation of a Clauser-Horne-Shimony-Holt \cite{Clauser:1969vv} (CHSH) type Bell inequality. To accomplish this, a CHSH $S$-value above the classical bound of $S\leq2$ needs to be experimentally obtained, being equivalent to an entanglement visibility of $1/\sqrt{2}\approx0.71$. This was only achievable when operating at low pump powers and, given the resulting low count rates, therefore only feasible in the course of a measurement performed locally on La Palma. We accumulated data over 8000 s and measured the required S-value for both the singlet $\ket{\Psi^-}_{03}$ and triplet $\ket{\Psi^+}_{03}$ state. In total we detected 5647 singlet and 5618 triplet swapping events and violated the inequality with $S_{singlet}=2.487\pm0.287$ and $S_{triplet}=2.469\pm0.287$ at a 15 kHz 2-fold rate, respectively (Figure \ref{figure3}). This result clearly proves that photons "0" and "3" have been projected into an entangled state.
\\
In order to reduce the accumulation time in the remote measurement scenario, we set each SPDC source to a local 2-fold rate of 130 kHz, corresponding to a locally detected 4-fold count rate of 100 Hz and an average entanglement visibility of the swapped state of approximately 0.60. We measured the expectation value of an entanglement-witness operator $W$, with $W < 0$ representing a sufficient condition for the presence of entanglement \cite{Guhne:2009kg}. Our entanglement-witness operator is given as
\begin{equation}
	W=\frac{1}{2}-\frac{1}{4}(1+V_{HV} +V_{PM} +V_{RL}),
\end{equation}
with $V_{HV}, V_{PM}, V_{RL}$ being the correlation visibilities of state $\ket{\Psi}_{03}$ in the three mutually unbiased bases horizontal/vertical (HV), plus/minus (PM) and right/left (RL), respectively. The visibility is given by $V=(CC_{max}-CC_{min})/(CC_{max}+CC_{min})$ with the max (min) coincidence counts $CC_{max}$ ($CC_{min}$). Inserting the measured visibilities into Eq.(3) yields a negative expectation value for the entanglement witness operator $W_{singlet}=-0.212\pm0.027$ and $W_{triplet}=-0.177\pm0.028$ with statistical significances of 7.99 and 6.37 standard deviations $\sigma$, respectively (assuming Poissonian photon statistics).  Hence, we unambiguously demonstrated that the experimentally obtained states between photon "0" and "3" have become entangled over 143 km (Figure \ref{figure3}). These results were obtained from subsequent 30 s data files, accumulated over a measurement time of 271 min during 4 consecutive nights. In total, 506 and 492 entanglement-swapping events have been recorded for the singlet and triplet state, respectively.

\subsection*{Concluding comments}
Our data demonstrate successful entanglement swapping via a long-distance free-space link under the influence of highly demanding environmental conditions, in fact more challenging than expected for a satellite-to-ground link. This proves the feasibility of quantum repeaters in a future space- and ground-based worldwide quantum internet.  In particular, in a quantum repeater scheme, a single step of the purification method realized in Ref.  \cite{Pan:2003kv} would increase our obtained visibilities beyond the bound for the violation of a CHSH type Bell inequality even in the remote scenario. Together with a reliable quantum memory, our results set the benchmark for an efficient quantum repeater at the heart of a global quantum-communication network.
\bigskip
\subsubsection*{Acknowledgements}
The authors thank X.-S. Ma for fruitful discussions and the staff of IAC: F. Sanchez-Martinez, A. Alonso, C. Warden, M. Serra, J. Carlos and the staff of ING: M. Balcells, C. Benn, J. Rey, O. Vaduvescu, A. Chopping, D. González, S. Rodríguez, M. Abreu, L. González, as well as J. Kuusela, E. Wille and Z. Sodnik from ESA for their support.
This work was made possible by grants from the European Space Agency (Contract 4000105798/12/NL/CBi), the Austrian Science Foundation (FWF) under projects SFB F4008 and CoQuS, the FFG within the ASAP 7 (No. 828316) program and the Federal Ministry of Science and Research (BMWF).
\bigskip
\subsubsection*{Author contribution}
T.H. conceived the research, designed, carried out the experiment and analyzed data with the help of T.S., M.F., J.H., B.W. and R.U.. A.Z. defined the scientific goals, conceived the research, designed and supervised the project and the experimental advancements. All authors contributed to the manuscript.
\bigskip


\begin{thebibliography}{10}
\expandafter\ifx\csname url\endcsname\relax
  \def\url#1{\texttt{#1}}\fi
\expandafter\ifx\csname urlprefix\endcsname\relax\def\urlprefix{URL }\fi
\providecommand{\bibinfo}[2]{#2}
\providecommand{\eprint}[2][]{\url{#2}}

\bibitem{Wootters:1982wj}
\bibinfo{author}{Wootters, W.~K.} \& \bibinfo{author}{Zurek, W.~H.}
\newblock \bibinfo{title}{{A single quantum cannot be cloned}}.
\newblock \emph{\bibinfo{journal}{Nature}} \textbf{\bibinfo{volume}{299}},
  \bibinfo{pages}{802--803} (\bibinfo{year}{1982}).

\bibitem{Bose:1998wq}
\bibinfo{author}{Bose, S.}, \bibinfo{author}{Vedral, V.} \&
  \bibinfo{author}{Knight, P.~L.}
\newblock \bibinfo{title}{{Multiparticle Generalization of Entanglement
  Swapping}}.
\newblock \emph{\bibinfo{journal}{Physical Review A}}
  \textbf{\bibinfo{volume}{57}}, \bibinfo{pages}{822--829}
  (\bibinfo{year}{1998}).

\bibitem{Kimble:2008if}
\bibinfo{author}{Kimble, H.~J.}
\newblock \bibinfo{title}{{The quantum internet}}.
\newblock \emph{\bibinfo{journal}{Nature}} \textbf{\bibinfo{volume}{453}},
  \bibinfo{pages}{1023--1030} (\bibinfo{year}{2008}).

\bibitem{Cirac:1999tz}
\bibinfo{author}{Cirac, J.~I.}, \bibinfo{author}{Ekert, A.~K.},
  \bibinfo{author}{Huelga, S.~F.} \& \bibinfo{author}{Macchiavello, C.}
\newblock \bibinfo{title}{{Distributed Quantum Computation over Noisy
  Channels}}.
\newblock \emph{\bibinfo{journal}{Physical Review A}}
  \textbf{\bibinfo{volume}{59}}, \bibinfo{pages}{4249--4254}
  (\bibinfo{year}{1999}).

\bibitem{Nielsen:2000vn}
\bibinfo{author}{Nielsen, M.~A.} \& \bibinfo{author}{Chuang, I.~L.}
\newblock \emph{\bibinfo{title}{{Quantum Computation and Quantum Information}}}
  (\bibinfo{publisher}{Cambridge University Press},
  \bibinfo{address}{Cambridge}, \bibinfo{year}{2000}).

\bibitem{Ladd:2010kq}
\bibinfo{author}{Ladd, T.~D.} \emph{et~al.}
\newblock \bibinfo{title}{{Quantum computers}}.
\newblock \emph{\bibinfo{journal}{Nature}} \textbf{\bibinfo{volume}{464}},
  \bibinfo{pages}{45--53} (\bibinfo{year}{2010}).

\bibitem{Zukowski:1993us}
\bibinfo{author}{Zukowski, M.}, \bibinfo{author}{Zeilinger, A.},
  \bibinfo{author}{Horne, M.~A.} \& \bibinfo{author}{Ekert, A.~K.}
\newblock \bibinfo{title}{{"Event-Ready-Detectors" Bell experiment via
  entanglement swapping}}.
\newblock \emph{\bibinfo{journal}{Physical Review Letters}}
  \textbf{\bibinfo{volume}{71}}, \bibinfo{pages}{4287--4290}
  (\bibinfo{year}{1993}).

\bibitem{Bennett:1996wz}
\bibinfo{author}{Bennett, C.~H.} \emph{et~al.}
\newblock \bibinfo{title}{{Purification of noisy entanglement and faithful
  teleportation via noisy channels}}.
\newblock \emph{\bibinfo{journal}{Physical Review Letters}}
  \textbf{\bibinfo{volume}{76}}, \bibinfo{pages}{722--725}
  (\bibinfo{year}{1996}).

\bibitem{Briegel:1998wp}
\bibinfo{author}{Briegel, H.~J.}, \bibinfo{author}{D{\"u}r, W.},
  \bibinfo{author}{Cirac, J.} \& \bibinfo{author}{Zoller, P.}
\newblock \bibinfo{title}{{Quantum repeaters: the role of imperfect local
  operations in quantum communication}}.
\newblock \emph{\bibinfo{journal}{Physical Review Letters}}
  \textbf{\bibinfo{volume}{81}}, \bibinfo{pages}{5932--5935}
  (\bibinfo{year}{1998}).

\bibitem{Pan:2001gg}
\bibinfo{author}{Pan, J.-W.}, \bibinfo{author}{Simon, C.},
  \bibinfo{author}{Brukner, C.} \& \bibinfo{author}{Zeilinger, A.}
\newblock \bibinfo{title}{{Entanglement purification for quantum
  communication}}.
\newblock \emph{\bibinfo{journal}{Nature}} \textbf{\bibinfo{volume}{410}},
  \bibinfo{pages}{1067--1070} (\bibinfo{year}{2001}).

\bibitem{Pan:2003kv}
\bibinfo{author}{Pan, J.-W.}, \bibinfo{author}{Gasparoni, S.},
  \bibinfo{author}{Ursin, R.}, \bibinfo{author}{Weihs, G.} \&
  \bibinfo{author}{Zeilinger, A.}
\newblock \bibinfo{title}{{Experimental entanglement purification of arbitrary
  unknown states}}.
\newblock \emph{\bibinfo{journal}{Nature}} \textbf{\bibinfo{volume}{423}},
  \bibinfo{pages}{417--422} (\bibinfo{year}{2003}).

\bibitem{Clausen:2011bw}
\bibinfo{author}{Clausen, C.} \emph{et~al.}
\newblock \bibinfo{title}{{Quantum storage of photonic entanglement in a
  crystal}}.
\newblock \emph{\bibinfo{journal}{Nature}} \textbf{\bibinfo{volume}{469}},
  \bibinfo{pages}{508--511} (\bibinfo{year}{2011}).

\bibitem{Gisin:2007vl}
\bibinfo{author}{Gisin, N.} \& \bibinfo{author}{Thew, R.}
\newblock \bibinfo{title}{{Quantum communication}}.
\newblock \emph{\bibinfo{journal}{Nature Photonics}}
  \textbf{\bibinfo{volume}{1}}, \bibinfo{pages}{165--171}
  (\bibinfo{year}{2007}).

\bibitem{Duan:2001tt}
\bibinfo{author}{Duan, L.~M.}, \bibinfo{author}{Lukin, M.~D.} \&
  \bibinfo{author}{Zoller, P.}
\newblock \bibinfo{title}{{Long-distance quantum communication with atomic
  ensembles and linear optics}}.
\newblock \emph{\bibinfo{journal}{Nature}} \textbf{\bibinfo{volume}{413}},
  \bibinfo{pages}{413--418} (\bibinfo{year}{2001}).

\bibitem{Barz:2012uw}
\bibinfo{author}{Barz, S.} \emph{et~al.}
\newblock \bibinfo{title}{{Demonstration of blind quantum computing}}.
\newblock \emph{\bibinfo{journal}{Science}} \textbf{\bibinfo{volume}{335}},
  \bibinfo{pages}{303--308} (\bibinfo{year}{2012}).

\bibitem{Pan:1998tn}
\bibinfo{author}{Pan, J.-W.}, \bibinfo{author}{Bouwmeester, D.},
  \bibinfo{author}{Weinfurter, H.} \& \bibinfo{author}{Zeilinger, A.}
\newblock \bibinfo{title}{{Experimental Entanglement Swapping: Entangling
  Photons That Never Interacted}}.
\newblock \emph{\bibinfo{journal}{Physical Review Letters}}
  \textbf{\bibinfo{volume}{80}}, \bibinfo{pages}{3891--3894}
  (\bibinfo{year}{1998}).

\bibitem{Jennewein:2001vs}
\bibinfo{author}{Jennewein, T.}, \bibinfo{author}{Weihs, G.},
  \bibinfo{author}{Pan, J.-W.} \& \bibinfo{author}{Zeilinger, A.}
\newblock \bibinfo{title}{{Experimental Nonlocality Proof of Quantum
  Teleportation and Entanglement Swapping}}.
\newblock \emph{\bibinfo{journal}{Physical Review Letters}}
  \textbf{\bibinfo{volume}{88}}, \bibinfo{pages}{17903} (\bibinfo{year}{2001}).

\bibitem{Halder:2007eo}
\bibinfo{author}{Halder, M.} \emph{et~al.}
\newblock \bibinfo{title}{{Entangling independent photons by time
  measurement}}.
\newblock \emph{\bibinfo{journal}{Nature Physics}}
  \textbf{\bibinfo{volume}{3}}, \bibinfo{pages}{692--695}
  (\bibinfo{year}{2007}).

\bibitem{Yuan:2008fj}
\bibinfo{author}{Yuan, Z.-S.} \emph{et~al.}
\newblock \bibinfo{title}{{Experimental demonstration of a BDCZ quantum
  repeater node}}.
\newblock \emph{\bibinfo{journal}{Nature}} \textbf{\bibinfo{volume}{454}},
  \bibinfo{pages}{1098--1101} (\bibinfo{year}{2008}).

\bibitem{Kaltenbaek:2009ix}
\bibinfo{author}{Kaltenbaek, R.}, \bibinfo{author}{Prevedel, R.},
  \bibinfo{author}{Aspelmeyer, M.} \& \bibinfo{author}{Zeilinger, A.}
\newblock \bibinfo{title}{{High-fidelity entanglement swapping with fully
  independent sources}}.
\newblock \emph{\bibinfo{journal}{Physical Review Letters}}
  \textbf{\bibinfo{volume}{79}}, \bibinfo{pages}{040302}
  (\bibinfo{year}{2009}).

\bibitem{Ma:2012kl}
\bibinfo{author}{Ma, X.-S.} \emph{et~al.}
\newblock \bibinfo{title}{{Experimental delayed-choice entanglement swapping}}.
\newblock \emph{\bibinfo{journal}{Nature Physics}}
  \textbf{\bibinfo{volume}{8}}, \bibinfo{pages}{479--484}
  (\bibinfo{year}{2012}).

\bibitem{Hofmann:2012jb}
\bibinfo{author}{Hofmann, J.} \emph{et~al.}
\newblock \bibinfo{title}{{Heralded Entanglement Between Widely Separated
  Atoms}}.
\newblock \emph{\bibinfo{journal}{Science}} \textbf{\bibinfo{volume}{337}},
  \bibinfo{pages}{72--75} (\bibinfo{year}{2012}).

\bibitem{Ekert:1991zz}
\bibinfo{author}{Ekert, A.~K.}
\newblock \bibinfo{title}{{Quantum cryptography based on Bell's theorem}}.
\newblock \emph{\bibinfo{journal}{Physical Review Letters}}
  \textbf{\bibinfo{volume}{67}}, \bibinfo{pages}{661--663}
  (\bibinfo{year}{1991}).

\bibitem{Scheidl:2010fj}
\bibinfo{author}{Scheidl, T.} \emph{et~al.}
\newblock \bibinfo{title}{{Violation of local realism with freedom of choice}}.
\newblock \emph{\bibinfo{journal}{Proceedings of the National Academy of
  Sciences}} \textbf{\bibinfo{volume}{107}}, \bibinfo{pages}{19708--19713}
  (\bibinfo{year}{2010}).

\bibitem{Calsamiglia:2001tl}
\bibinfo{author}{Calsamiglia, J.} \& \bibinfo{author}{L{\"u}tkenhaus, N.}
\newblock \bibinfo{title}{{Maximum Efficiency of a Linear-Optical Bell-State
  Analyzer}}.
\newblock \emph{\bibinfo{journal}{Applied Physics B}}
  \textbf{\bibinfo{volume}{72}}, \bibinfo{pages}{67--71}
  (\bibinfo{year}{2001}).

\bibitem{Kwiat:1995ub}
\bibinfo{author}{Kwiat, P.~G.} \emph{et~al.}
\newblock \bibinfo{title}{{New High-Intensity Source of Polarization-Entangled
  Photon Pairs}}.
\newblock \emph{\bibinfo{journal}{Physical Review Letters}}
  \textbf{\bibinfo{volume}{75}}, \bibinfo{pages}{4337--4342}
  (\bibinfo{year}{1995}).

\bibitem{Kim:2002gd}
\bibinfo{author}{Kim, Y.-H.} \& \bibinfo{author}{Grice, W.~P.}
\newblock \bibinfo{title}{{Generation of pulsed polarization-entangled
  two-photon state via temporal and spectral engineering}}.
\newblock \emph{\bibinfo{journal}{Journal of Modern Optics}}
  \textbf{\bibinfo{volume}{49}}, \bibinfo{pages}{2309--2323}
  (\bibinfo{year}{2002}).

\bibitem{Kim:2003tva}
\bibinfo{author}{Kim, Y.-H.}, \bibinfo{author}{Kulik, S.~P.},
  \bibinfo{author}{Chekhova, M.~V.}, \bibinfo{author}{Grice, W.~P.} \&
  \bibinfo{author}{Shih, Y.}
\newblock \bibinfo{title}{{Experimental entanglement concentration and
  universal Bell-state synthesizer}}.
\newblock \emph{\bibinfo{journal}{Physical Review A}}
  \textbf{\bibinfo{volume}{67}}, \bibinfo{pages}{010301}
  (\bibinfo{year}{2003}).

\bibitem{Poh:2009iq}
\bibinfo{author}{Poh, H.}, \bibinfo{author}{Lim, J.},
  \bibinfo{author}{Marcikic, I.}, \bibinfo{author}{Lamas-Linares, A.} \&
  \bibinfo{author}{Kurtsiefer, C.}
\newblock \bibinfo{title}{{Eliminating spectral distinguishability in ultrafast
  spontaneous parametric down-conversion}}.
\newblock \emph{\bibinfo{journal}{Physical Review Letters}}
  \textbf{\bibinfo{volume}{80}}, \bibinfo{pages}{043815}
  (\bibinfo{year}{2009}).

\bibitem{Hong:1987vi}
\bibinfo{author}{Hong, C.~K.}, \bibinfo{author}{Ou, Z.~Y.} \&
  \bibinfo{author}{Mandel, L.}
\newblock \bibinfo{title}{{Measurement of subpicosecond time intervals between
  two photons by interference.}}
\newblock \emph{\bibinfo{journal}{Physical Review Letters}}
  \textbf{\bibinfo{volume}{59}}, \bibinfo{pages}{2044--2046}
  (\bibinfo{year}{1987}).

\bibitem{Ma:2012ei}
\bibinfo{author}{Ma, X.-S.} \emph{et~al.}
\newblock \bibinfo{title}{{Quantum teleportation over 143 kilometres using
  active feed-forward}}.
\newblock \emph{\bibinfo{journal}{Nature}} \textbf{\bibinfo{volume}{489}},
  \bibinfo{pages}{269--273} (\bibinfo{year}{2012}).

\bibitem{Clauser:1969vv}
\bibinfo{author}{Clauser, J.~F.}, \bibinfo{author}{Horne, M.~A.},
  \bibinfo{author}{Shimony, A.} \& \bibinfo{author}{Holt, R.~A.}
\newblock \bibinfo{title}{{Proposed Experiment to Test Local Hidden-Variable
  Theories}}.
\newblock \emph{\bibinfo{journal}{Physical Review Letters}}
  \textbf{\bibinfo{volume}{23}}, \bibinfo{pages}{880--884}
  (\bibinfo{year}{1969}).

\bibitem{Guhne:2009kg}
\bibinfo{author}{G{\"u}hne, O.} \& \bibinfo{author}{T{\'o}th, G.}
\newblock \bibinfo{title}{{Entanglement detection}}.
\newblock \emph{\bibinfo{journal}{Physics Reports}}
  \textbf{\bibinfo{volume}{474}}, \bibinfo{pages}{1--75}
  (\bibinfo{year}{2009}).

\end{thebibliography}

\end{document}